\documentclass[12pt]{article}

\usepackage{setspace}

\usepackage[utf8]{inputenc}
\usepackage{amssymb,amsmath,bm,amssymb,physics,mathtools,amsthm}
\usepackage[flushleft]{threeparttable}
\usepackage{dcolumn}
\usepackage[titletoc]{appendix}
\usepackage{color}
\usepackage{nameref}
\usepackage[usenames,dvipsnames,svgnames,table]{xcolor}
\usepackage[colorlinks=true]{hyperref}
\usepackage{pgfplots} 

\usepackage{natbib}
\hypersetup{colorlinks=true,allcolors=[rgb]{0,0,0.45}}
\usepackage{adjustbox}
\usepackage{booktabs} 
\usepackage{array} 
\usepackage{paralist}
\usepackage{diagbox}
\usepackage{threeparttable}

\usepackage{float}
\usepackage{upgreek}
\usepackage{bbm}
\usepackage[outdir=./]{epstopdf}

\newtheorem{theorem}{Theorem}
\newtheorem{proposition}{Proposition}

\theoremstyle{definition}
\newtheorem{definition}{Definition}

\usepackage{geometry} 
\usepackage{graphicx} 
\usepackage{booktabs} 
\usepackage{array} 
\usepackage{paralist}
\usepackage{verbatim}
\usepackage{subfig}
\usepackage{fancyhdr}
\usepackage{sectsty}
\usepackage[nottoc,notlof,notlot]{tocbibind} 
\usepackage[titles,subfigure]{tocloft}

\title{Exponential Discounting under Partial Efficiency} 

\author{Charles Gauthier\thanks{KU Leuven, Leuven, Belgium (email: charles.gauthier@kuleuven.be). I wish to thank Victor Aguiar for his helpful comments and suggestions. I also wish to thank Roy Allen, Audra Bowlus, Maria Goltsman, Nail Kashaev, David Rivers, and anonymous referees for useful comments.}}
\date{This version: May 2024}

\begin{document}
	\maketitle
	
	\begin{abstract}
	\noindent
        This paper derives a novel representation of the exponential discounting model that allows one to assess departures from the model via a measure of efficiency. The approach uses a revealed preference methodology that does not make any parametric assumption on the utility function and allows for unrestricted heterogeneity. The method is illustrated using longitudinal data from checkout scanners and gives insights into sources of departure from exponential discounting.
		
		
	\end{abstract}

	\section{Introduction} \label{Section1}
	
        The exponential discounting model is a predominant tool for analyzing dynamic choice in applied work. Its attractiveness rests in that time preferences are summarized by a single parameter\textemdash the discount factor. This allows one to tractably analyze a decision maker's intertemporal choices, which is crucial in a vast range of applications. However, a common feature in this literature is the specification of the consumer preferences. This constitutes a potentially important limitation as erroneously specifying preferences may lead to the erroneous rejection of the model.
	
        At its core, the exponential discounting model assumes that the utility function is additively time-separable and stationary. Under these assumptions, the transitivity of preferences can be characterized by the well-known Generalized Axiom of Revealed Preference (GARP). In particular, \cite{Afriat1967} showed that for any finite data set $\left\lbrace (\bm{\rho}_t, \bm{c}_t) \right\rbrace_{t \in \mathcal{T}}$ of discounted prices and demands, GARP is necessary and sufficient for the existence of a well-behaved utility function that rationalizes the data. The distinctive feature of exponential discounting, though, is the prediction that consumers will be time consistent. Namely, it requires consumers to commit to their initial plan as time unfolds.\footnote{In experimental settings, a preference reversal occurs when the consumer chooses a sooner-smaller reward over a later-larger one and then switches to the later-larger reward when an equal delay is added to both outcomes. This behavior violates time consistency if the consumer deviates from his plan and chooses the sooner-smaller reward in the future \citep{Halevy2015}.}
	
    This paper shows that the exponential discounting model, which is normally stated as a dynamic maximization problem with an intertemporal budget constraint, may be expressed as a repeated static utility maximization problem without a budget constraint. Using this novel representation, I propose an efficiency index similar to the critical cost efficiency index (CCEI) of \hyperlink{Afriat}{Afriat $(1973)$} to account for small optimization errors in the exponential discounting model. Importantly, this index can be decomposed in two mutually exclusive indices based on a similar approach as \cite{HH2019}. Precisely, this new index can be split into an index that captures deviations from GARP and another that captures deviations from time consistency. As such, I am able to gain new insights on the sources of departure from exponential discounting.
 
    The novel representation of exponential discounting is equivalent to that of \cite{Browning1989} but can easily accommodate optimization errors. A different strand of literature extends exponential discounting to richer settings such as preference heterogeneity and renegotiations within the household \citep{Adams2014} and habit formation \citep{Crawford2010,DV2013}. In a different direction, \cite{EIS2021} provide an axiomatic characterization of exponential discounting that applies in experimental data sets.\footnote{That is, their approach applies to a situation where consumers make multiple intertemporal choices over a single good. They also provide a characterization of other models such as quasi-hyperbolic discounting.} The current approach rather applies in field data where consumers make choices over multidimensional bundles and focuses on understanding deviations from the model.

	\section{Exponential Discounting Model} \label{Section3}

	\subsection{Notation}
	
	The typical scenario under consideration is that of purchases made by a consumer over a finite time window. Let $\mathcal{L} \in \{1, ...,L\}$ denote the number of observed commodities and $\mathcal{T} = \{0, ..., T\}$ the periods for which data on consumers are observable. For any good $l \in \mathcal{L}$ and time period $t \in \mathcal{T}$, denote discounted price by $\rho_{l,t} = p_{l,t}/ \prod_{i=0}^{t} (1 + r_i)$, where $p_{l,t}$ is the spot price and $r_i$ is the interest rate, and denote consumption by $c_{l,t}$.\footnote{The interest rate in the first period is set to zero, that is, $r_0 = 0$.} An observation is therefore a pair $(\bm{\rho}_t, \bm{c}_t) \in \mathbb{R}_{++}^{L} \times \mathbb{R}_{+}^{L}$ and, accordingly, a data set is written as $\left\lbrace ( \bm{\rho}_{t},\bm{c}_{t}) \right\rbrace_{t \in \mathcal{T}}$.
	
	\subsection{Exponential Discounting Rationalizability}
	
	The objective function faced by an exponential discounting (ED) consumer at time $\tau \in \mathcal{T}$ is given by
	\begin{equation*}
	U_{\tau}(\bm{c}_{\tau}, ..., \bm{c}_{T-\tau}) = u(\bm{c}_{\tau}) + \sum_{j=1}^{T - \tau} \delta^{j} u(\bm{c}_{\tau + j}),
	\end{equation*}
	\noindent
	where $u(\cdot)$ is the instantaneous utility function and $\delta \in (0,1]$ is the discount factor. Moreover, consumption satisfies the linear budget constraint
	\begin{equation*}
	\boldsymbol{\rho}'_t \bm{c}_t + s_{t}^{d} = y_{t}^{d} + a_{t}^{d} \quad \forall t \in \{\tau, \dots, T\},
	\end{equation*}
	
	\noindent
	where $s_{t}^{d}$ denotes discounted savings, $y_{t}^{d} > 0$ denotes discounted income and $a_t$ is the discounted value of assets held at period $t$.\footnote{That is, $s_{t}^{d} = s_{t}/ \prod_{i=0}^{t} (1 + r_i)$, $y_{t}^{d} = y_{t}/ \prod_{i=0}^{t} (1 + r_i)$ and $a_{t}^{d} = a_{t}/ \prod_{i=0}^{t} (1 + r_i)$.} The assets evolve according to the law of motion: $a_t = (1 + r_t) s_{t-1}$. A data set is consistent with exponential discounting if it can be thought of as stemming from the model. 
	\begin{definition}
	    A data set $\left\lbrace ( \bm{\rho}_{t},\bm{c}_{t}) \right\rbrace_{t \in \mathcal{T}}$ is ED-rationalizable if there exist a locally nonsatiated, continuous, monotonic, and concave instantaneous utility function $u(\cdot)$, an income stream $(y_{t}^{d})_{t \in \mathcal{T}} \in \mathbb{R}_{++}^{|\mathcal{T}|}$, an initial asset level $a_0 \geq 0$, and a discount factor $\delta \in (0,1]$ such that the consumption stream $(\bm{c}_t)_{t \in \mathcal{T}}$ solves
		\begin{equation*}
			\underset{ (\bm{c}_t)_{t \in \mathcal{T}} \in \mathbb{R}_{+}^{L \times |\mathcal{T}|}}{\max} \; u(\bm{c}_{0}) + \sum_{t=1}^{T} \delta^{t} u(\bm{c}_t) \; \; \text{s.t.} \; \; \boldsymbol{\rho}'_0 \bm{c}_0 + \sum_{t=1}^{T} \boldsymbol{\rho}'_t \bm{c}_t  = y_0 + \sum_{t=1}^{T} y_{t}^{d} + a_0.
		\end{equation*}
	\end{definition}
	
	Consistent with the permanent income hypothesis, I assume that the marginal utility of discounted expenditure is constant across time \citep{Bewley1977, Browning1989}. This is motivated by the fact that, if the marginal utility of discounted expenditure at $t$ was higher than at $s \neq t$, then the consumer could move income from $s$ to $t$ such as to increase his consumption at $t$ and be better off. The following result provides a novel representation of the exponential discounting model.

	\begin{theorem} \label{theorem1}
		The following statements are equivalent:
		\renewcommand{\theenumi}{\roman{enumi}}
		\begin{enumerate}
			\item[{(i)}] The data set $\left\lbrace ( \bm{\rho}_{t},\bm{c}_{t}) \right\rbrace_{t \in \mathcal{T}}$ is ED-rationalizable.

			\item[{(ii)}] There exist a locally nonsatiated, continuous, monotonic and concave instantaneous utility function $u(\cdot)$ and a discount factor $\delta \in (0,1]$ such that for all $t \in \mathcal{T}$ and $\bm{c} \in \mathbb{R}_{+}^{L}$
			\begin{equation*}
			    u(\bm{c}_t) - \delta^{-t} \boldsymbol{\rho}'_t \bm{c}_t \geq u(\bm{c}) - \delta^{-t} \boldsymbol{\rho}'_t \bm{c}.
			\end{equation*}
		\end{enumerate}
	\end{theorem}

    \noindent
	The representation in Theorem \ref{theorem1} $(ii)$ has two distinctive features. First, there is no budget constraint. Second, the consumer problem is much simpler as it only requires solving for optimal consumption bundles rather than the whole consumption stream. To interpret the condition in Theorem \ref{theorem1} $(ii)$, it is useful to rewrite it as
	\begin{equation*}
	\bm{c}_t \in \underset{\bm{c} \in \mathbb{R}_{+}^{L}}{\text{arg} \max} \; u(\bm{c}) + \delta^{-t}(y_{t}^{d} - \bm{\rho}'_t \bm{c}) \quad \forall t \in \mathcal{T}.
	\end{equation*}
	
	This formulation emphasizes that exponential discounting can be seen as a repeated static utility maximization problem. Letting $s^{d} := y_{t}^{d} - \bm{\rho}'_t \bm{c}$ denote savings and $U(\bm{c}, s^{d}) :=  u(\bm{c}) + \delta^{-t} s^{d}$, the objective function can be interpreted as an augmented utility function $U : \mathbb{R}_{+}^{L} \times \mathbb{R} \rightarrow \mathbb{R}$.\footnote{See \cite{Deb2023} for a different use of an augmented utility function.} It indicates that, in any given time period, the consumer values both current consumption and savings. This compromise between current consumption and savings captures the idea that increasing consumption today leaves a lesser amount of wealth for future periods, thus diminishing future consumption. In the absence of a budget constraint, the mechanism by which an interior solution is achieved therefore relies on the trade-off between the two.

	\subsection{Exponential Discounting Rationalizability under Partial Efficiency}
	 
    Using the novel representation of the exponential discounting model, I follow the revealed preference literature and relax its constraints with an efficiency level $e \in (0,1]$.
	\begin{definition}
		Let $e \in (0,1]$. The $e$-ED model rationalizes the data $\left\lbrace ( \bm{\rho}_{t},\bm{c}_{t}) \right\rbrace_{t \in \mathcal{T}}$ if there exist a locally nonsatiated, continuous, monotonic and concave utility function $u(\cdot)$ and a discount factor $\delta \in (0,1]$ such that for all $t \in \mathcal{T}$ and $\bm{c} \in \mathbb{R}_{+}^{L}$
		\begin{equation*}
		    u(\bm{c}_t) - \delta^{-t} \boldsymbol{\rho}'_t \bm{c}_t \geq u(\bm{c}) - \delta^{-t} \boldsymbol{\rho}'_t \bm{c}/e.
		\end{equation*}
	\end{definition}
	
	This definition accounts for digressions from exponential discounting by considering an efficiency level $e$ that rationalizes every choice of a consumer at once.\footnote{Alternatively, one could have an efficiency index for each choice as in \cite{Varian1990}, and then consider some aggregator function \citep{Dziewulski2020} to determine the overall level of inefficiency.} In particular, note that any consumption behavior may be rationalized by the $e$-ED model for an $e$ arbitrarily close to zero. The following result extends the analysis of exponential discounting to account for optimization errors.
	\begin{proposition} \label{lemmaprime}
		For a given $e \in (0,1]$, the following statements are equivalent:
		\renewcommand{\theenumi}{\roman{enumi}}
		\begin{enumerate}
			\item[{(i)}] There exist a locally nonsatiated, continuous, monotonic and concave utility function 
			$u(\cdot)$ and a discount factor $\delta \in (0,1]$ $e$-ED rationalizing the data $\left\lbrace ( \bm{\rho}_{t},\bm{c}_{t}) \right\rbrace_{t \in \mathcal{T}}$. 
			
			\item[{(ii)}] There exist numbers $u_{t}$, $t = 0, \dots, T,$ and a discount factor $\delta \in (0,1]$ such that
			\begin{equation*}
			u_s \leq u_{t} + \delta^{-t} \boldsymbol{\rho}'_t (\bm{c}_s /e - \bm{c}_{t}) \quad \forall s,t \in \mathcal{T}.
			\end{equation*}
		\end{enumerate}
	\end{proposition}

	Conditional on $(e,\delta) \in (0,1]^{2}$, the existence of a solution can be checked by solving the set of inequalities in Proposition \ref{lemmaprime} $(ii)$ using linear programming. One can recover nonparametric bounds on the discount factor by finding the set of all discount factors consistent with these inequalities at a given $e \in (0,1]$. Note that a data set that needs a small efficiency level to be $e$-ED rationalizable is farther away from exponential discounting than one with a large efficiency level. In particular, if $e = 1$ then the data set is ED-rationalizable.

    \subsection{Exponential Efficiency Index}
    
    As with the critical cost efficiency index (CCEI) proposed by \cite{Afriat1973}, one can consider the largest efficiency level rationalizing the data for the exponential discounting model. Formally, I define the exponential efficiency index as
    \begin{equation*}
	   \text{EEI} := \text{sup} \{e \in [0,1]: \left\lbrace ( \bm{\rho}_{t},\bm{c}_{t}) \right\rbrace_{t \in \mathcal{T}} \; \text{is} \; \text{$e$-ED rationalizable}\}.
    \end{equation*}

    The EEI provides a measure of distance between a data set and the exponential discounting model. However, it does not differentiate between deviations arising from within-period consistency (GARP) and time consistency.\footnote{The Generalized Axiom of Revealed Preference (GARP) is equivalent to the model of static utility maximization and is implied by exponential discounting \citep{Browning1989}.} To disentangle sources of departure from exponential discounting, an efficiency measure that controls for violations of static utility maximization is needed. I call such an efficiency measure the time consistency efficiency index (TCEI). 
    
    The TCEI can be derived based on the $2$-step rationalization procedure of \cite{HH2019} for homothetic rationalizability. The first stage consists in finding the largest efficiency level rationalizing the data with respect to static utility maximization and yields
    \begin{equation*}
	   u_s \leq u_t + \lambda_t \boldsymbol{\rho}'_{t} (\bm{c}_s/\footnotesize{\scriptsize{\text{CCEI}}} - \bm{c}_t) \quad \forall s,t \in \mathcal{T}.
    \end{equation*}

\noindent
Imposing the additional restriction of the exponential discounting model to the CCEI-Afriat inequalities amounts to setting $\lambda_t = \delta^{-t}$ and yields
\begin{equation*}
	u_s \leq u_t + \delta^{-t} \boldsymbol{\rho}'_{t} (\bm{c}_s/ \scriptsize{\text{CCEI}} - \bm{c}_t) \quad \forall s,t \in \mathcal{T}.\vspace{-0.4cm}
\end{equation*}
	
\noindent
The TCEI then corresponds to the largest efficiency level rationalizing the previous system of inequalities with respect to the $e$-ED model:
\begin{equation*}
	u_s \leq u_t + \delta^{-t} \boldsymbol{\rho}'_{t} \left(\frac{\bm{c}_s}{\scriptsize{\text{CCEI} \cdot \text{TCEI}}} - \bm{c}_t\right) \quad \forall s,t \in \mathcal{T}.
\end{equation*}

\noindent
Since the largest efficiency level that solves the $e$-ED model is the EEI, it follows that EEI $=$ CCEI $\cdot$ TCEI. One can therefore recover the TCEI by first obtaining the CCEI and the EEI.

	\section{Empirical Application} \label{Section6}
	
	\subsection{Data}
	
	In my empirical analysis, I implement the methodology developed in the previous sections using the Stanford Basket Dataset, which is a panel data set containing expenditures of $494$ households between June $1991$ and June $1993$.\footnote{I treat households as unitary entities even though they may have many members. As such, I refer to a household as an individual.} Specifically, I use the transformed data set of \cite{ELS2011}. As such, goods for which prices are observed in every week are retained and aggregated by brand for periods of four weeks. This yields a total of $375$ distinct goods belonging to one of the following $14$ categories: bacon, barbecue sauce, butter, cereal, coffee, crackers, eggs, ice cream, nuts, analgesics, pizza, snacks, sugar and yogurt. As the data do not contain information on interest rates, I include interest rates on personal loans at commercial banks from the Federal Reserve Bank of St. Louis.\footnote{Since the data on interest rates are quarterly, I use a linear interpolation to obtain monthly observations.} For a comprehensive description of the scanner data set, I refer the reader to \cite{ELS2011}.

    \subsection{Results}
    
    In what follows, I first compute the efficiency indices for static utility maximization (CCEI), time consistency (TCEI), and exponential discounting (EEI). I let $\mu_e$ denote the mean, $\sigma_e$ denote the standard deviation, $\underline{e}$ denote the smallest value of the efficiency index, and $\overline{e}$ denote the largest value off the efficiency index. The results presented in Table \ref{tab:3} are obtained with a grid search over $\delta \in (0,1]$ with a step size of $0.01$ and a binary search algorithm for the efficiency indices that guarantees them to be within $2^{-10}$ of their true values.
    \begin{center}
    \captionof{table}{Rationalizability Results} \label{tab:3}
	\begin{adjustbox}{width=9cm,center=\textwidth}
		\begin{tabular}{cc|cccccccccc}
			\toprule
			\vspace{-0.5cm} \\
			\multicolumn{1}{c}{Efficiency index} & \multicolumn{1}{c|}{N}  &  $\underline{e}$       &      $\overline{e}$         &     $\mu_{e}$      &  $\sigma_{e}$     \\
			\midrule  
			CCEI & 494 & 0.6865   &  1.0000   &  0.9551 & 0.0502  \\
			\midrule 
			TCEI & 494 & 0.4758   &  1.0000   &  0.8365 & 0.0802 \\
			\midrule
			EEI & 494 & 0.3878      & 0.9561  & 0.7984 & 0.0820 \\
			\bottomrule
		\end{tabular}
	\end{adjustbox}
	\caption*{\small \textit{Notes:} The sample size is N $= 494$. $\underline{e}$ denotes the lowest efficiency index, $\overline{e}$ the largest efficiency index, $\mu_{e}$ the average efficiency index, and $\sigma_{e}$ the standard deviation of the efficiency index.}
\end{center}
    	
The results in Table \ref{tab:3} indicate that time consistency is a more stringent assumption than GARP, with an average efficiency level for the TCEI below that of the CCEI by approximately $0.10$.

Next, I use the EEI to recover nonparametric bounds on each individual discount factor. The greatest lower bound and least upper bound on the discount factor define the identified set ($IS)$. Summary results are presented in Table \ref{tab:1stat} using a grid search over $\delta \in (0,1]$ with a step size of $0.001$. For ease of exposition, summary statistics on the discount factor use the midpoint of the identified set. 
\begin{center}
	\captionof{table}{Nonparametric Bounds on Discount Factors} \label{tab:1stat} 
	\begin{adjustbox}{width=1 \textwidth,center=\textwidth}
		\begin{tabular}{ c c| c c c c c c c c} 
			\toprule
			\multicolumn{1}{c}{Efficiency index} & \multicolumn{1}{c|}{N} & \multicolumn{0}{c}{$\underline{IS}$} & \multicolumn{0}{c}{$\overline{IS}$} &   \multicolumn{0}{c}{$\mu_{IS}$} & \multicolumn{0}{c}{$\sigma_{IS}$} & \multicolumn{0}{c}{$\underline{\delta}$} & \multicolumn{0}{c}{$\overline{\delta}$} & \multicolumn{0}{c}{$\mu_{\delta}$} & \multicolumn{0}{c}{$\sigma_{\delta}$} \\
			\hline
			EEI & $494$ & 0.000 & 0.013 & 0.005 & 0.003 & 0.837 & 0.924 & 0.877 & 0.007\\
			\bottomrule
		\end{tabular}
	\end{adjustbox}
	\caption*{\small \textit{Notes:} The sample size is N $= 494$. $\underline{IS}$ denote the smallest size of the identified set, $\overline{IS}$ the largest size of the identified set, $\mu_{IS}$ the average size of the identified set, and $\sigma_{IS}$ the standard deviation of the identified set's size. Discount factors are defined as the midpoint of the identified set. $\underline{\delta}$ denotes the lowest discount factor, $\overline{\delta}$ the largest discount factor, $\mu_{\delta}$ the average discount factor, and $\sigma_{\delta}$ the standard deviation of the discount factor.}
\end{center}

 The first four columns of Table \ref{tab:1stat} show that bounds on the discount factor are very informative in spite of their nonparametric nature. The last four columns further show that the discount factor, defined as the midpoint of the identified set, is about $0.877$ on average in the data.

\section{Conclusion} \label{Section7}

In this paper, I recognize that deviations from exponential discounting may naturally arise due to small optimization or measurement errors. My results allow one to assess the extent of departure from the exponential discounting model and to determine if utility maximization or time consistency is at fault. More broadly, the revealed preference inequalities derived in this study could be used to bound a consumer's response to a change in prices. The methodology could therefore be used to do robust welfare analysis, estimate market power in empirical industrial organization, or determine optimal pricing schemes in marketing.\\

	\noindent
	\hypertarget{I}{{\Large\textbf{Appendix}}}\\
	
	\noindent
	{\large\textbf{Proof of Theorem 1}}
	
	\noindent

	\noindent
	$(i) \implies (ii)$ \hypertarget{theorem(i)}{}
	
	\noindent
	From the first-order condition, we have
	\begin{equation*}
	\nabla u(\bm{c}_t) \leq \delta^{-t} \boldsymbol{\rho}_{t} \quad \forall t \in \mathcal{T},
	\end{equation*}
	
	\noindent
	where $\nabla u(\bm{c}_t)$ is some supergradient of $u(\cdot)$ at $\bm{c}_t$. By continuity and concavity of the instantaneous utility function, we know that for all $t \in \mathcal{T}$ and $\bm{c} \in \mathbb{R}_{+}^{L}$
	\begin{equation*}
	u(\bm{c})\leq u(\bm{c}_t)+\nabla u(\bm{c}_t)' (\bm{c} - \bm{c}_t).
	\end{equation*}
	
	Let $N$ be a set of indices such that $\nabla u(\bm{c}_t)_j = \delta^{-t} \boldsymbol{\rho}_{t,j}$ for all $j \in N$. It follows that $\nabla u(\bm{c}_t)_j \leq \delta^{-t} \boldsymbol{\rho}_{t,j}$ for all $j \notin N$. Thus, $\bm{c}_{t,j} = 0$ is a corner solution for all $j \notin N$. We therefore have
	\begin{align*}
	u(\bm{c}) - u(\bm{c}_t) \leq \nabla u(\bm{c}_t)' (\bm{c} - \bm{c}_t) &= \sum_{j \in N} \nabla u(\bm{c}_t)_j (\bm{c}_{j} - \bm{c}_{t,j}) + \sum_{j \notin N} \nabla u(\bm{c}_t)_j (\bm{c}_{j} - \bm{c}_{t,j}) \\
	& = \sum_{j \in N} \delta^{-t} \boldsymbol{\rho}_{t,j} (\bm{c}_{j} - \bm{c}_{t,j}) + \sum_{j \notin N} \nabla u(\bm{c}_t)_j (\bm{c}_{j} - \bm{c}_{t,j}) \\
	& \leq \sum_{j \in N} \delta^{-t} \boldsymbol{\rho}_{t,j} (\bm{c}_{j} - \bm{c}_{t,j}) +\sum_{j \notin N} \delta^{-t} \boldsymbol{\rho}_{t,j} (\bm{c}_{j} - \bm{c}_{t,j}),
	\end{align*}
	
	\noindent
	where the last inequality holds since $\bm{c}_{t,j} = 0$ and $\bm{c}_{j} \geq 0$ for all $j \notin N$. As a result, for all $t \in \mathcal{T}$ and $\bm{c} \in \mathbb{R}_{+}^{L}$
	\begin{equation*}
	u(\bm{c})\leq u(\bm{c}_t) + \delta^{-t} \bm{\rho}'_{t}(\bm{c} - \bm{c}_t).
	\end{equation*}
	
	\noindent
	Rearranging gives that for all $t \in \mathcal{T}$ and $\bm{c} \in \mathbb{R}_{+}^{L}$
	\begin{equation*}
	u(\bm{c}_t) - \delta^{-t} \bm{\rho}'_t \bm{c}_t \geq  u(\bm{c}) - \delta^{-t} \bm{\rho}'_{t}\bm{c},
	\end{equation*}
	
	\noindent
	where, by assumption, the instantaneous utility function is locally nonsatiated, continuous, monotonic, and concave and $\delta \in (0,1]$.\\
	
		\noindent
	$(ii) \implies (i)$ \hypertarget{theorem(ii)}{}
	
	\noindent
	The instantaneous utility function is locally nonsatiated, continuous, monotonic, and concave and the discount factor satisfies $\delta \in (0,1]$. For all $t \in \mathcal{T}$ and $\bm{c} \in \mathbb{R}_{+}^{L}$, we also have
	\begin{equation*}
	    u(\bm{c}_t) - \delta^{-t} \boldsymbol{\rho}'_t \bm{c}_t \geq u(\bm{c}) - \delta^{-t} \boldsymbol{\rho}'_t \bm{c}.
	\end{equation*}
	
	\noindent
	Rearranging gives that for all $t \in \mathcal{T}$ and $\bm{c} \in \mathbb{R}_{+}^{L}$
	\begin{equation*}
	u(\bm{c}) \leq u(\bm{c}_t) + \delta^{-t} \boldsymbol{\rho}'_t (\bm{c} - \bm{c}_t).
	\end{equation*}
	
	\noindent
	This inequality corresponds to the definition of concavity and, therefore, it follows that $\delta^{-t} \boldsymbol{\rho}_t$ is a supergradient of $u(\cdot)$ at $\bm{c}_t$ for all $t \in \mathcal{T}$. \\

	\noindent
	{\large\textbf{Proof of Proposition $1$}}
	
	\noindent
	$(i) \implies (ii)$
	
	\noindent
	Since the data set $\left\lbrace ( \bm{\rho}_{t},\bm{c}_{t}) \right\rbrace_{t \in \mathcal{T}}$ is $e$-ED rationalizable, it is the case that for all $t \in \mathcal{T}$ and $\bm{c} \in \mathbb{R}_{+}^{L}$
	\begin{equation*}
	    u(\bm{c}_t) - \delta^{-t} \boldsymbol{\rho}'_t \bm{c}_t \geq u(\bm{c}) - \delta^{-t} \boldsymbol{\rho}'_t \bm{c}/e
	\end{equation*}

    \noindent	
	for some $\delta \in (0,1]$. By the same argument as in the proof of Theorem $1$ \hyperlink{theorem(ii)}{(ii)}, we can obtain
	\begin{equation*}
	u(\bm{c}_s) \leq u(\bold{c_t}) + \delta^{-t} \boldsymbol{\rho}'_t (\bm{c}_s/e - \bm{c}_t) \quad \forall s, t \in \mathcal{T}, 
	\end{equation*}
	
	\noindent
	where one may define $u_t := u(\bm{c}_t)$ for all $t \in \mathcal{T}$.\\
 
	\noindent
	$(ii) \implies (i)$
	
	\noindent
    Consider any sequence of indices $\tau = \{t_i\}_{i =1}^{m}$, $t_i \in \mathcal{T}$, $m \geq 2$, and let $\mathcal{I}$ be the set of all such indices. Summing up the inequalities for the resulting cycle yields
	\begin{equation*}
	0 \leq \delta^{-t_1} \boldsymbol{\rho}'_{t_1} (\bm{c}_{t_2}/e - \bm{c}_{t_1}) + \ldots + \delta^{-t_m} \boldsymbol{\rho}'_{t_m} (\bm{c}_{t_1}/e - \bm{c}_{t_m}).
	\end{equation*}

    \noindent
	For some $e \in (0,1]$, define
	\begin{equation*}
	u(\bm{c}) = \underset{\tau \in \mathcal{I}}{\text{inf}} \Big\{ \delta^{-\tau(m)} \boldsymbol{\rho}'_{\tau(m)} \big(\bm{c}/e - \bm{c}_{\tau(m)}\big) + \sum_{i=1}^{m-1} \delta^{- \tau(i)} \boldsymbol{\rho}'_{\tau(i)} \big(\bm{c}_{\tau(i+1)}/e - \bm{c}_{\tau(i)}\big) \Big\}.
	\end{equation*}
	
	\noindent
	This utility function is locally nonsatiated, continuous, monotonic and concave as it is the pointwise minimum of a collection of affine functions. Moreover, the infimum defining $u(\bm{c})$ has no cycle of indices. Consider $\bm{c} \in \mathbb{R}_{+}^{L}$ such that $\bm{c} \neq \bm{c}_t$ and let $\tau_t \in \mathcal{I}$ be a minimizing sequence for $\bm{c}_t$. It follows that
	\begin{align*}
	u(\bm{c}) - \delta^{-t} \boldsymbol{\rho}'_t \bm{c}/e &\leq \delta^{-t} \boldsymbol{\rho}'_{t} \big(\bm{c}/e - \bm{c}_t \big) + \delta^{- \tau_t (m_t)} \boldsymbol{\rho}'_{\tau_t (m_t)} \big(\bm{c}_t/e - \bm{c}_{\tau_t (m_t)}\big) \\ & \quad + \sum_{i=1}^{m_t - 1} \delta^{-\tau_t(i)} \boldsymbol{\rho}'_{\tau_t(i)} \big(\bm{c}_{\tau_t(i+1)}/e - \bm{c}_{\tau_t(i)}\big) - \delta^{-t} \boldsymbol{\rho}'_t \bm{c}/e \\
	&=  \delta^{- \tau_t (m_t)} \boldsymbol{\rho}'_{\tau_t (m_t)} \big(\bm{c}_t/e - \bm{c}_{\tau_t (m_t)}\big) \\& \quad + \sum_{i=1}^{m_t - 1} \delta^{-\tau_t(i)} \boldsymbol{\rho}'_{\tau_t(i)} \big(\bm{c}_{\tau_t(i+1)}/e - \bm{c}_{\tau_t(i)}\big) - \delta^{-t} \boldsymbol{\rho}'_t \bm{c}_t \\
	&= u(\bm{c}_t) - \delta^{-t} \boldsymbol{\rho}'_t \bm{c}_t,
	\end{align*}
	
	\noindent
	where the first inequality holds since $u(\bm{c})$ uses the sequence achieving the infimum for $\bm{c}$, the first equality is a mere simplification, and the last equality is a consequence of $\tau_t$ being a minimizing sequence for $\bm{c}_t$. Thus, I have shown the existence of a locally nonsatiated, continuous, monotonic and concave utility function and a discount factor $\delta \in (0,1]$ that $e$-ED rationalize the data.

    \section{Statements and Declarations}
    
    The author did not receive support from any organization for the submitted work. The author has no relevant financial or non-financial interests to disclose. The dataset used in this study is available in the Mendeley Data repository: \url{https://data.mendeley.com/datasets/9js2bj7c33/2}. DOI: 10.17632/9js2bj7c33.2.

    \bibliographystyle{econ1}
    \bibliography{references_new}

\end{document}